# Wetting of ferrofluids: phenomena and control

Mika Latikka[a,*], Matilda Backholm[a], Jaakko V. I. Timonen[a], Robin H. A. Ras[a,b]
[a] Department of Applied Physics, Aalto University School of Science
P.O. Box 15100, FI-00076 Aalto, Espoo, Finland
[b] Department of Bioproducts and Biosystems, Aalto University School of Chemical Engineering,
P.O. Box 16100, FI-00076 Aalto, Espoo, Finland
[*] Corresponding author

E-mail addresses: mika.latikka@aalto.fi, matilda.backholm@aalto.fi, jaakko.timonen@aalto.fi, robin.ras@aalto.fi

Keywords: ferrofluid, wetting, magnetic field, contact angle, magnetic nanoparticles

## Abstract

Ferrofluids are liquids exhibiting remarkably strong response to magnetic fields, which leads to fascinating properties useful in various applications. Understanding the wetting properties and spreading of ferrofluids is important for their use in microfluidics and magnetic actuation. However, this is challenging as magnetically induced deformation of the ferrofluid surface can affect contact angles, which are commonly used to characterize wetting properties in other systems. In addition, interaction of the magnetic nanoparticles and solid surface at nanoscale can have surprising effects on ferrofluid spreading. In this review we discuss these issues with focus on interpretation of ferrofluid contact angles. We review recent literature examining ferrofluid wetting phenomena and outline novel wetting related ferrofluid applications. To better understand wetting of ferrofluids, more careful experimental work is needed.

## 1. Introduction

Ferrofluids are colloidal suspensions of small (~3-15 nm diameter[1]) superparamagnetic nanoparticles in a liquid carrier medium. They combine liquid properties with a strong magnetic response, and thus exhibit fascinating phenomena, such as field-induced pattern formation on the ferrofluid surface[2], self-assembly of droplets[3] and magnetic micro-convection[4] (Fig 1). The superparamagnetic nanoparticles used in ferrofluids usually consist of ferri- or ferromagnetic metals or metal oxides [5*]. Due to the small nanoparticle size, Brownian motion is enough to keep the particles from settling in gravitational or magnetic fields [2]. The small size also renders the nanoparticles superparamagnetic: each particle acts as a single magnetic domain (although surface effects and structural defects can influence this [6]), which can flip its magnetization direction due to thermal agitation. Due to this Neél relaxation together with rotational Brownian relaxation, ferrofluids show no remanent magnetization in room temperature once the external magnetic field is removed [5*]. This lack of magnetic hysteresis is also exhibited by paramagnetic materials, but otherwise superparamagnetic particles resemble ferri- or ferromagnetic materials with high susceptibility and nonlinear response to magnetic fields [1].

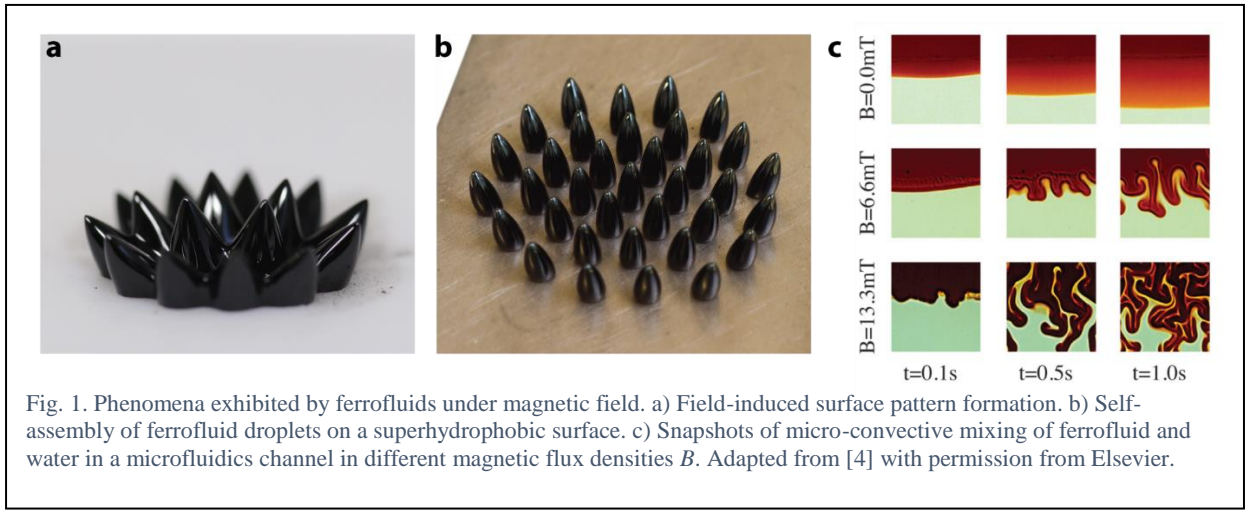

Fig. 1. Phenomena exhibited by ferrofluids under magnetic field. a) Field-induced surface pattern formation. b) Self-assembly of ferrofluid droplets on a superhydrophobic surface. c) Snapshots of micro-convective mixing of ferrofluid and water in a microfluidics channel in different magnetic flux densities $B$. Adapted from [4] with permission from Elsevier.

Superparamagnetic nanoparticles can be fabricated using either a top-down approach, such as grinding of larger particles [7], or a bottom-up approach, i.e. chemical synthesis. The latter is nowadays more widely used, especially chemical co-precipitation involving $Fe^{3+}$ and $Fe^{2+}$ salts in water [5*], [8]. To prevent aggregation caused by attractive forces between the particles, such as van der Waals and magnetic dipole interactions [9], the nanoparticles need to be stabilized. This is done by introducing interparticle repulsion by charging the particles (electrostatic stabilization) or by coating them with capping agents (steric stabilization) [5*]. These capping agents are usually surfactants, especially in case of ferrofluids with organic carrier liquid. For a more thorough introduction to the properties of ferrofluids we direct the reader to an excellent recent review by Torres-Díaza and Rinaldi [10**].

Stable, magnetically controllable liquids are useful in a range of mechanical to biomedical applications. Ferrofluids are widely used as liquid seals and lubricants held in place by magnetic fields, while actuation with dynamic magnetic fields allow building of ferrofluid based pumps, valves and tunable optical systems [5*], [10**]. Their anisotropic heat transfer capabilities find use as heat transfer fluids and magnetic buoyancy can be exploited in separation processes [5*], [10**]. Recently ferrofluids have been increasingly investigated for microfluidic [11*]–[13*] and biomedical [14*] applications. In many of these cases, the ferrofluid wetting properties need to be carefully understood and tuned in order to ensure reliable function of these systems. In this review we discuss the wetting properties of ferrofluids, how they can be controlled with external magnetic fields and some interesting applications from the past few years.

## 2. Wetting of ferrofluids

### 2.1 Contact angles

Wetting characterization relies heavily on the concept of contact angle, which is the angle between the solid-liquid and liquid-air interfaces [15*]. The smaller the angle, the better the liquid wets the surface. The contact angle value typically depends on the length scale it is measured on (Fig 2 a-c) [16]–[19]. The microscopic, actual contact angle existing locally at each point on the contact line would be the ideal quantity for measuring wetting properties, but it is unfortunately very difficult to probe experimentally (Fig 2 b). Usually a macroscopic apparent contact angle (measured at length scales over tens of micrometres) is used instead, which represents an average value of the microscopic local contact angles (Fig 2 a). Care must be taken when measuring the apparent contact angle, because body forces (for example gravity) not related to wetting properties can deform the droplet profile. Because of this, the apparent contact angle must be determined using appropriate magnification. Finally, all three phases (solid, liquid and gas) are very close to one another near the contact line at the nanoscale. This gives rise to a disjoining pressure, which determines the curvature of the liquid surface near (<100 nm) the solid surface (Fig 2 c) [16]–[18]. The exact surface geometry is again difficult to probe experimentally due to the small scale.

The apparent contact angle can be interpreted to reflect the average molecular wetting properties, that is the interfacial tensions between solid, liquid and surrounding fluid (immiscible liquid or gas), which is why it is used for wetting characterization. This angle is rarely unique, but can instead have a range of metastable values, for example due to surface inhomogeneities. The highest metastable value is called advancing contact angle, which appears when the wetting front is slowly advancing on previously dry surface (Fig 2 d). Similarly, the lowest metastable value, receding contact angle, is seen when the wetting front is slowly receding from previously wetted surface (Fig 2 e). This range of metastable states, called contact angle hysteresis, is related to the mobility of the contact line. If the contact line is strongly pinned, the droplet can be deformed with large forces before it starts to move. Droplet mobility can also be assessed by measuring the roll-off angle (the angle of inclination needed to unpin the droplet from the surface). It is determined by depositing a droplet on a surface and tilting it until the droplet starts to move.

Because of the contact angle hysteresis, a difference in contact angles does not automatically imply a difference in wetting properties. For the same reason, reporting a single value of contact angle (often called a static contact angle) gives only limited information about wetting properties of the system. Furthermore, the macroscopic droplet profile depends also on body forces acting on the liquid, such as gravity, which can affect the measurement of apparent contact angles.

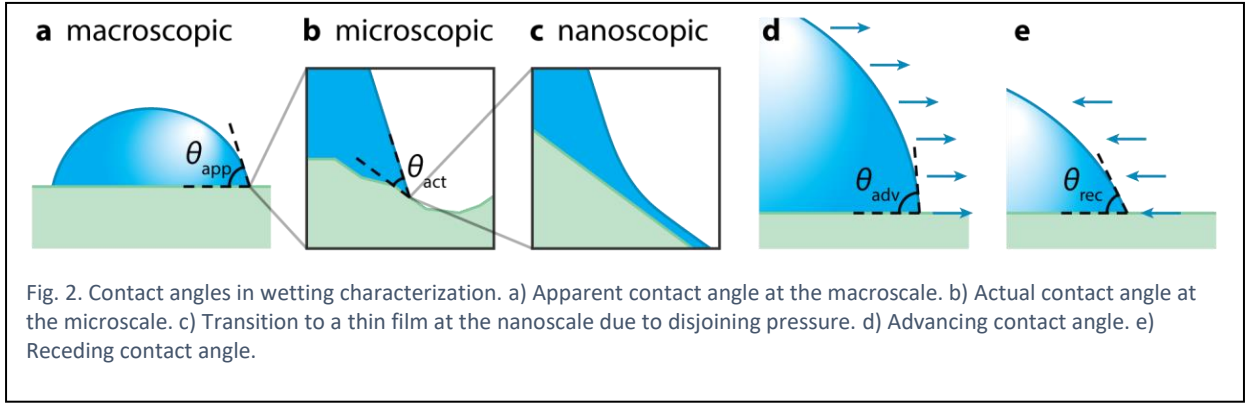

Fig. 2. Contact angles in wetting characterization. a) Apparent contact angle at the macroscale. b) Actual contact angle at the microscale. c) Transition to a thin film at the nanoscale due to disjoining pressure. d) Advancing contact angle. e) Receding contact angle.

## 2.2 Ferrofluid shape in magnetic field

Arguably the most interesting quality of ferrofluids is the possibility to control the liquid flow and shape of the liquid surface with a magnetic field. This arises from the interactions between magnetic field and the dipole moments of each nanoparticle. The related forces can be described with a magnetic stress tensor **T** [2]:

$$T_{ij} = -\left\{\int_0^H \mu_0 \left[\frac{\partial(Mv)}{\partial v}\right]_{H,T} dH + \frac{1}{2}\mu_0 H^2\right\}\delta_{ij} + B_i H_j \quad (1)$$

where $H$ is the magnetic field intensity, $\mu_0$ is the vacuum permeability, $M$ is the fluid magnetization, $v$ is the specific volume ($v = \rho^{-1}$, $\rho$ is the density), $T$ is the temperature, $\delta_{ij}$ is the Kronecker delta function and $B$ is the magnetic induction. Here $H$ and $B$ refer to local total field quantities, not to applied external fields. The subject is complicated and only the most important results can be shown here. For a more complete description and derivations see *Ferrohydrodynamics* by Rosensweig [2] or a more recent treatment by Stierstadt and Liu [20].

Magnetic body-force density can be calculated from Eq. (1) as:

$$\mathbf{f}_m = \nabla \cdot \mathbf{T} = -\nabla\left\{\mu_0 \int_0^H \left[\frac{\partial(Mv)}{\partial v}\right]_{H,T} dH\right\} + \mu_0 M \nabla H \quad (2)$$

The integral term vanishes for dilute ferrofluids, but for concentrated ferrofluids it cannot always be neglected, because particle density can affect dipolar interactions between the nanoparticles. The last term proportional to field gradient $\nabla H$ describes a force pulling the magnetic material toward higher field strength. For example, a sessile ferrofluid droplet flattens when a permanent magnet with a strong field gradient is placed underneath the surface supporting the droplet. This force can also be used, for example, to move a ferrofluid droplet or to manipulate non-magnetic objects immersed in ferrofluid [21].

The steady-state flow of inviscid, incompressible ferrofluid can be described by the ferrohydrodynamic Bernoulli equation [2]:

$$p^* + \frac{1}{2}\rho v^2 + \rho g h - p_m = \text{const} \quad (3)$$

$$p^* = p + \mu_0 \int_0^H \left[\frac{\partial(Mv)}{\partial v}\right]_{H,T} dH \tag{4}$$

$$p_m = \mu_0 \int_0^H M dH \tag{5}$$

where $p^*$ is the composite pressure in the magnetic fluid, $p$ is the thermodynamic pressure, $p_m$ is the fluid-magnetic pressure, $v$ is velocity, $g$ is the acceleration of gravity and $h$ is the elevation from a chosen reference level. In addition, the following boundary condition must be met [2]:

$$p^* + p_n = p_0 + p_c \tag{6}$$

$$p_n = \frac{\mu_0}{2} M_n^2 \tag{7}$$

$$p_c = K\gamma \tag{8}$$

where $p_n$ is the magnetic normal traction due to field continuity requirements at the interface, $p_0$ is the pressure outside the magnetic fluid, $p_c$ is the capillary pressure, $M_n$ is the fluid magnetization component normal to the fluid surface and $K$ is the sum of principal surface curvatures.

As a simple example, let us consider a deep pool of ferrofluid that is partly exposed to a local vertical uniform magnetic field created by an electromagnet. Far away from the magnet the field is zero and the Bernoulli equation on the fluid surface simplifies to $p_1^* + \rho g h_1 = $ const. On the fluid surface at the axis of the magnet, the equation is $p_2^* + \rho g h_2 - \mu_0 \int_0^H M dH = $ const. The surface is assumed flat. Boundary conditions from Eq. (6) give $p_1^* = p_0$ and $p_2^* = p_0 - \frac{\mu_0}{2} M^2$ ($M = M_n$ in a field normal to the ferrofluid surface). The ferrofluid in magnetic field rises compared to the ferrofluid in zero field [2]:

$$\Delta h = h_2 - h_1 = \frac{1}{\rho g}\left(\mu_0 \int_0^H M dH + \frac{\mu_0}{2} M^2\right) \tag{9}$$

Similarly a ferrofluid droplet in uniform magnetic field elongates along the field direction. This deformation increases the surface area, and thus also the energy due to interfacial tension $E_\gamma = \int_A \gamma dA$, which was neglected from the previous example. The equilibrium droplet shape is then determined by balancing the gravitational, magnetic and surface energies. In case of small droplets gravity can be ignored and the relative strengths of uniform magnetic field and capillarity can be conveniently described with a magnetic Bond number $N_m = \mu_0 H^2 R/\gamma$, where $R$ is the drop diameter and $\gamma$ is the surface tension [22].

Fig 3 a shows a silicone oil-based ferrofluid droplet suspended in glycerol in uniform magnetic fields from 6.032 to 162.33 kA/m [23]. Calculating the exact shape is not easy even in case of a uniform magnetic field, as droplet magnetization affects the total magnetic field $H$ [24]. An example of a magnetized sphere in uniform magnetic field is given in Fig. 3 b. A change in droplet shape further alters the magnetic field geometry, making the calculation of the magnetic energy $E_m = \int_V \int_0^B \mathbf{H} \cdot d\mathbf{B}\, dV$ and the equilibrium droplet shape difficult [2].

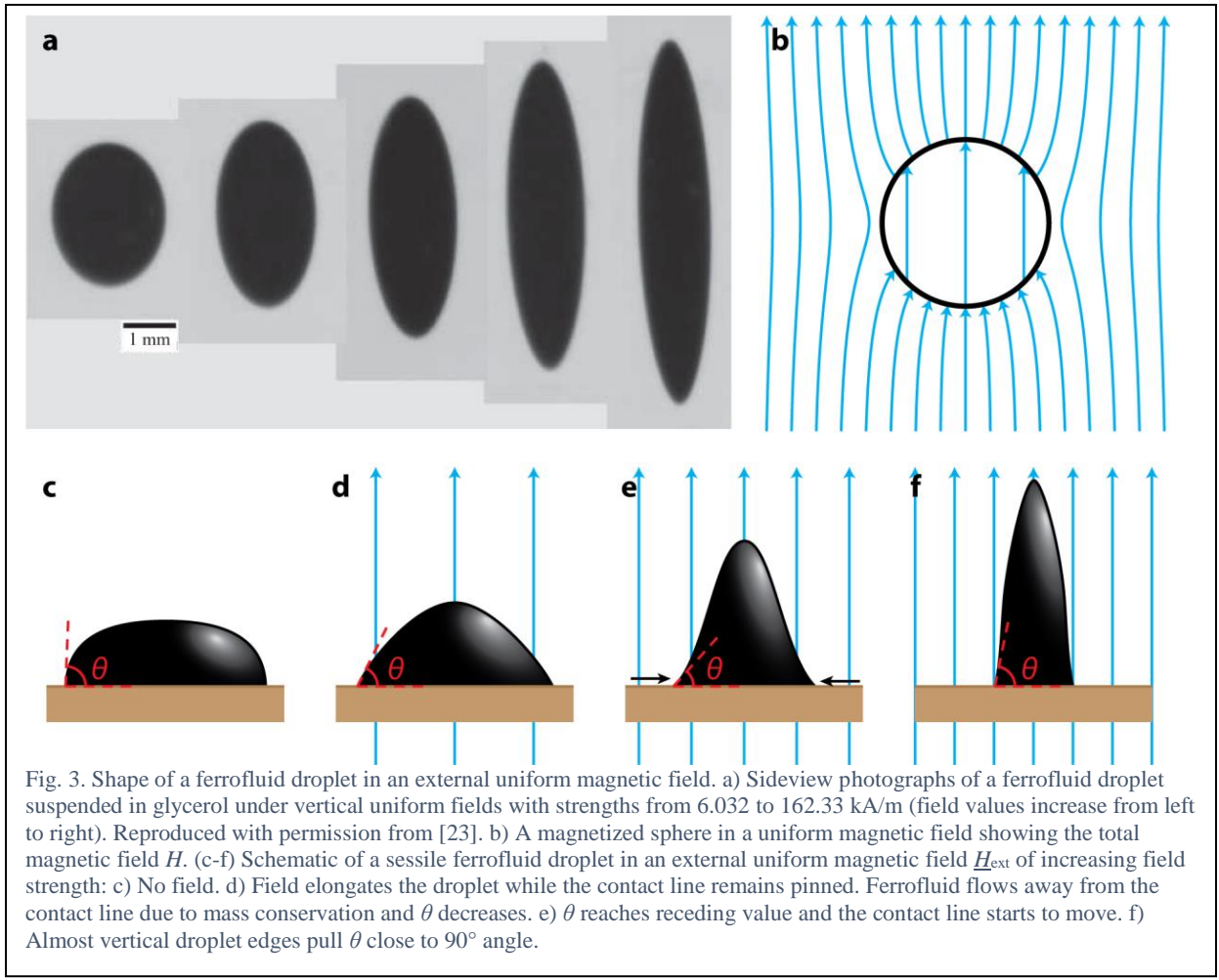

Fig. 3. Shape of a ferrofluid droplet in an external uniform magnetic field. a) Sideview photographs of a ferrofluid droplet suspended in glycerol under vertical uniform fields with strengths from 6.032 to 162.33 kA/m (field values increase from left to right). Reproduced with permission from [23]. b) A magnetized sphere in a uniform magnetic field showing the total magnetic field $H$. (c-f) Schematic of a sessile ferrofluid droplet in an external uniform magnetic field $H_{ext}$ of increasing field strength: c) No field. d) Field elongates the droplet while the contact line remains pinned. Ferrofluid flows away from the contact line due to mass conservation and $\theta$ decreases. e) $\theta$ reaches receding value and the contact line starts to move. f) Almost vertical droplet edges pull $\theta$ close to 90° angle.

## 2.3 Effects of magnetic field on interfacial tension

Droplet shapes and wetting phenomena are governed by minimization of surface energies within given boundary conditions. Unfortunately, interfacial tensions under magnetic fields have not been thoroughly investigated. Experimental studies with ferrofluids are especially lacking and usually the interfacial tensions are assumed to be independent of magnetic fields. However, magnetic fields can have a large effect on tension values, for example in case of magnetically responsive surfactants [25].

Afkhami *et al.* compared simulated ellipsoidal shape of a ferrofluid droplet suspended in immiscible fluid to experimental and analytical results [23]. While the agreement was good in weak uniform magnetic fields, there were apparent deviations between numerical and experimental droplet shapes in fields above 12 kA/m. Authors were able to solve this discrepancy by allowing the interfacial tension of the droplet to change with the magnetic field in the simulations. This apparent change of several mN/m in interfacial tension was speculated to arise from rearrangement of nanoparticles within the fluid and at the interface. However, Rowghanian *et al.* were able to reproduce the experimental data in their theoretical description with a single interfacial tension value [24]. According to them the discrepancy between the theoretical prediction and experimental results in [23] was due to the ellipsoidal shape approximation of a ferrofluid droplet.

Kalikmanov proposed a statistical theory, where ferrofluid carrier liquid molecules and magnetic dipoles are treated separately [26]. It was shown that direct magnetic contribution to the surface tension was negligible but that dipole interactions can cause a non-uniform nanoparticle distribution, and as a result surface tension increases weakly with magnetic field strength. A more complex model using surface tension tensor and surface magnetization was recently developed by Zhukov [27]. The model predicts a relative change of 0.5% in the interfacial tension tangential component in 14 kA/m magnetic field parallel to the interface of water and a suspension with 15 vol-% of magnetic nanoparticles in dodecane. According to the results, tangential magnetic field increases the interfacial tension, whereas normal field decreases it. Slight increase in ferrofluid surface tension in external magnetic field has also been experimentally observed [28]. Field-induced change in interfacial tensions affects also the contact angles [29].

## 2.4 Effects of magnetic field on contact angles

Care must be taken when assessing wetting properties under external fields using contact angle goniometry. While magnetic fields can affect the interfacial tensions, and thus contact angles, this effect can be difficult to observe because of the deformation of the whole ferrofluid surface. As pointed out in section 2.1, contact angles reflect the forces acting on the liquid interface. These forces vary depending on the length scale under investigation, and not all of them are related to the wetting properties of the system. In other words, an external field can distort the interface and the apparent contact angle without affecting any intrinsic wetting properties.

The change in apparent contact angle due to external field has been studied extensively in the case of electric fields. This phenomenon is called electrowetting and it was originally interpreted to originate from a change in solid-liquid interfacial tension due to the electric field [30]. However, the decrease in apparent contact angle and spreading of the liquid can be explained, arguably better, with electrostatic pressure that deforms the profile of the liquid surface [31], [32]. As a consequence, at the microscopic scale the contact angle equals Young contact angle, which has since been shown also experimentally [33]. Similarly, it has been long hypothesized that while the apparent contact angle changes with magnetic field, the microscopic contact angle remains the same [34]. There are also some experimental results that support this view [22]. It is important to notice that the magnetic field induced curvature of a ferrofluid surface can be much higher compared to gravity, since magnetic forces on ferrofluid can be several orders of magnitude stronger than gravitational force. This means that higher magnification might be needed when measuring ferrofluid contact angles in magnetic fields than in mere gravitational field.

The apparent effect of magnetic field on ferrofluid contact angles was reported already in 1983 [35]. When a sessile ferrofluid droplet was elongated with a magnetic field, the contact angle was observed to decrease. However, as pointed out in the section 2.1, a change in apparent contact angle does not necessarily mean that the wetting properties have changed. The contact angle could have remained within the contact angle hysteresis range, reflecting only a change in droplet shape. In more recent experiments, flattening of a ferrofluid droplet accompanied by a decrease in contact angle from approximately 70° to 50° was observed in the field created by a permanent magnet under the substrate [36]. Due to the field gradient, the droplet experiences a downwards force and spreads on the surface, as described in section 2.2. The magnet was then moved laterally and the ferrofluid droplet followed with advancing and receding contact angles of 64° and 46°, respectively. Interestingly the reported contact angle in absence of a magnetic field was higher than the advancing contact angle in a magnetic field, even though the advancing contact angle is supposed to be the highest metastable angle. This could be due to insufficient magnification while measuring the contact angles or indeed suggests an actual change in intrinsic wetting properties due to the magnetic field.

Recently Rigoni *et al*. investigated ferrofluid droplet shape and contact angle dependence on field parameters and nanoparticle concentration [37*]. Depending on the magnetization and field gradient the droplets were observed to either elongate or flatten in magnetic field created by a permanent magnet. The contact angle (approximately 110°) hardly varied while the droplets were flattened by increasing the magnetic field. For elongated droplets the contact angle on the other hand decreased to 60° as the droplet height increased in the field. However, during droplet deformation the contact line remained pinned, and thus the change in contact angle probably reflects the change in droplet shape rather than wetting properties.

Droplet shape hysteresis was recently experimentally studied with ferrofluid ($Fe_3O_4$ nanoparticles with mean diameter of 10 nm in light hydrocarbon carrier liquid) and magnetic paint ($Fe_3O_4$-coated flake pigments with mean diameter of approximately 35 µm in water) [38]. In the experiments the magnetic field was first increased and then decreased by controlling the separation between the droplet and a permanent magnet underneath it. The ferrofluid droplet was observed to elongate when the field was increased from 0 to 20 mT. On the other hand, the magnetic paint droplet flattened when the field was changed from 0 to 330 mT. This conflicting behavior was explained to originate from different magnetic properties of the carrier fluids, even though both liquids contained large amounts of ferro- or superparamagnetic particles, as evident from their strong response to magnetic fields. A more likely explanation is that the microparticles in the magnetic paint sedimented because of the magnetic field. Curiously, the contact angle of the ferrofluid droplet first decreased from 45° to 25° and then increased to 38° as a function of the magnetic field. The initial decrease can be easily understood due to mass conservation: fluid is flowing from the contact line to the elongating peak, resulting in a decrease in contact angle. The following increase in angle is accompanied by decreasing contact area, and arises when the droplet surface away from the peak becomes more and more vertical, thus pulling the contact angle closer to 90° angle. Again, it can be argued that the change in contact angle due to magnetic field does not reflect a change in liquid-solid wetting properties, but happens within the range of normal contact angle hysteresis and is due to changing droplet shape. Similar schematic example of a ferrofluid droplet in an increasing uniform field is shown in Figure 3: first the liquid flows away from contact line to the axis of the droplet, reducing contact angle (c and d). Contact line starts to move when the angle is equal to receding contact angle (e). Finally, the angle increases again as the droplet surface becomes more vertical due to the normal stresses induced by the magnetic field (f).

Magnetically induced change in ferrofluid contact angle was recently investigated also numerically using the Young-Laplace equation and by taking into account the normal stresses created by the magnetic field [39]. Downward magnetic force flattened the droplet and decreased the contact angle, whereas upward magnetic force elongated the droplet and increased the contact angle. It is important to notice that these numerical results correspond to approximate equilibrium contact angles, and cannot be directly compared to experimental results, where contact angle hysteresis plays a role. Complex dependency of contact angles on magnetic field was found for a cylindrical sessile nanodroplet of Ising fluid using density functional theory [40]. On one hand magnetization increased the attraction between magnetic fluid molecules, which lead to an increase in contact angle. On the other hand, in a non-uniform field created by a permanent magnet under the droplet, the fluid was pulled towards the solid surface deforming the droplet and decreasing the contact angle. Even though the model relied on Ising interactions instead of dipole interactions, it agreed qualitatively with experiments performed with ferrofluids.

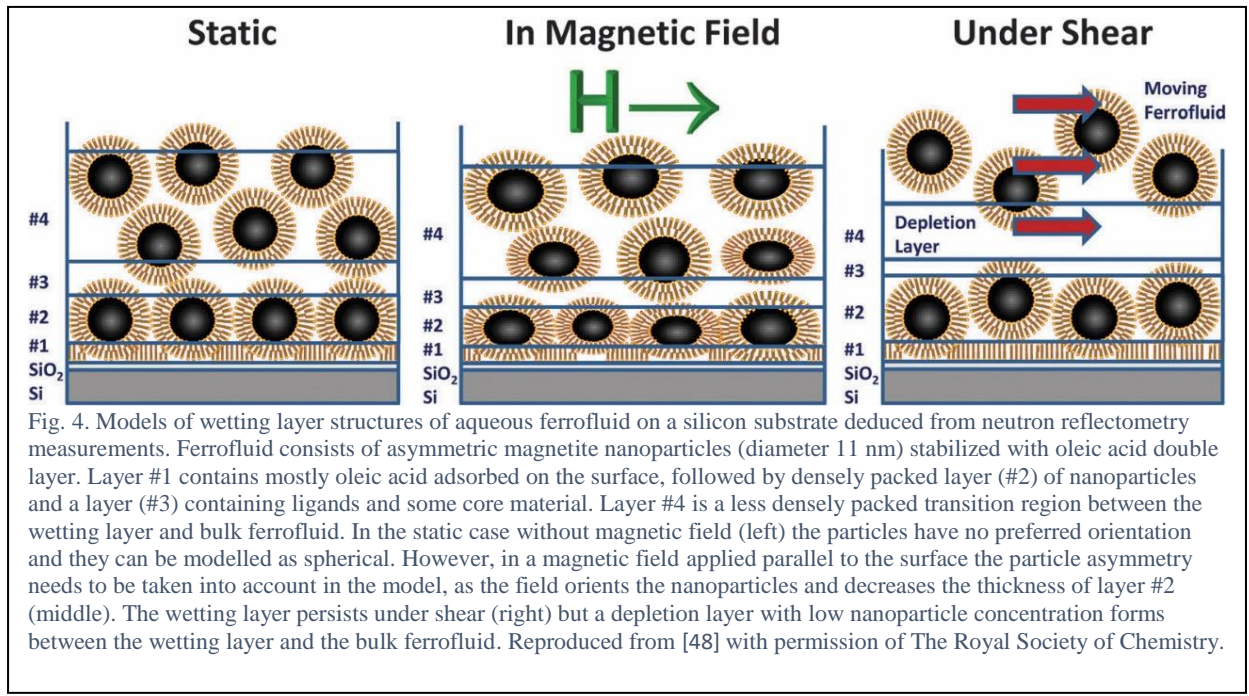

Fig. 4. Models of wetting layer structures of aqueous ferrofluid on a silicon substrate deduced from neutron reflectometry measurements. Ferrofluid consists of asymmetric magnetite nanoparticles (diameter 11 nm) stabilized with oleic acid double layer. Layer #1 contains mostly oleic acid adsorbed on the surface, followed by densely packed layer (#2) of nanoparticles and a layer (#3) containing ligands and some core material. Layer #4 is a less densely packed transition region between the wetting layer and bulk ferrofluid. In the static case without magnetic field (left) the particles have no preferred orientation and they can be modelled as spherical. However, in a magnetic field applied parallel to the surface the particle asymmetry needs to be taken into account in the model, as the field orients the nanoparticles and decreases the thickness of layer #2 (middle). The wetting layer persists under shear (right) but a depletion layer with low nanoparticle concentration forms between the wetting layer and the bulk ferrofluid. Reproduced from [48] with permission of The Royal Society of Chemistry.

## 2.5 Nanoparticle adhesion

Ferrofluids are complex liquids consisting of carrier fluid and dispersed colloidal nanoparticles. Both components play an important role in the wetting properties of these fluids. The carrier liquid typically contains capping agents used to stabilize the nanoparticles, such as surfactants, which can dramatically alter the surface tension and wetting properties of the ferrofluid compared to a pure carrier liquid. Surfactants not only change the surface tension of the ferrofluid, but can affect the wetting properties also by adsorbing to solid-liquid interface. This surfactant layer can remain on the surface when the ferrofluid retracts from the substrate, changing its surface free energy. As a result, the wetting properties can be different depending on whether the surface has been previously wetted by the ferrofluid or not, which needs to be taken into account when investigating ferrofluid wetting. It is also important to remember that surfactant diffusion to the interfaces is a dynamical process, and changing the interfacial area can also change the interfacial tension [41]. Because of this, dynamic measurements or long equilibration times are often needed.

The nanoparticles themselves can also affect the spreading of ferrofluid in a complex way: the particles tend to adsorb on the solid surface and near the contact line, typically enhancing spreading, as discussed by Wasan *et al*. in their review [42]. This phenomenon can have counter-intuitive effects. For example, an increase in nanoparticle concentration increases the viscosity of the fluid, which should decrease the wetting velocity. On the contrary, however, the velocity was observed to increase due to ordering of nanoparticles near the wetting front. The arising microstructures created a disjoining pressure gradient on the film near the contact line, which enhanced the wetting process.

The adsorption of colloidal nanoparticles on surfaces can be caused by London-van der Waals and electrical double layer forces, described by DLVO theory, or solvation, hydrophobic and steric forces [43]. Ordering of magnetic nanoparticles on surfaces has been investigated both theoretically [44] and experimentally. Magalhães *et al*. explored nematic assembly of ionic ferrofluids on untreated and PTFE-coated glass by measuring birefringence of the fluid near the surface [45], [46]. On the other hand, smectic-like ordering was found in ferrofluid near $SiO_2$ surface using in-situ neutron reflectometry [47]. The nanoparticles formed a wetting double-layer on the hydrophilic surface, which could be extended to 15 colloidal layers with a moderate uniform magnetic field perpendicular to the surface. A field parallel to the surface, however, lead to only short-ranged and perturbed ordering. Similar results were recently reported by Theis-Bröhl *et al*. using magnetite nanoparticles stabilized by carboxylic acid (Fig 4) [48]. They found that carboxylic acid covers the $SiO_2$ surface eliminating most of the water from the wetting layer created by the nanoparticles. Under shear this layer stayed adsorbed on the surface while a depletion layer was formed between the wetting layer and flowing bulk ferrofluid. With a magnetic field parallel to surface, slow reorientation of the particles was observed in the wetting layer along the field.

As with other nanofluids, ferrofluids show complex, time-dependent wetting properties even without the application of an external field. When a magnetic field is applied, a variety of additional phenomena emerges.

## 3. Field-induced control

### 3.1 Spreading of ferrofluids

While it can be argued that magnetic fields only slightly change intrinsic wetting properties between ferrofluid and solid, they can alter the force balance on the contact line, affecting ferrofluid spreading. A classic example is a ferrofluid film rising against gravity on a vertical current carrying wire due to the magnetic field induced around the conductor [49]. When the current is increased over a certain threshold value, the ferrofluid film height is suddenly increased. This has been treated as analogous to the wetting transition caused by van der Waals forces [50]. The same phenomenon with different boundary conditions has since been studied extensively using more sophisticated numerical models, also very recently [51], [52]. Similar magnetic field induced changes in spreading have also been investigated experimentally on vertical walls [53], nanochannels [54], porous media [55], and in the cases of thin films [56], [57] and droplets [35]–[38].

Ferrofluid spreading depends strongly on the geometry of the magnetic field. For example, Rosensweig *et al*. investigated a vertical wall partially immersed in a pool of ferrofluid in uniform magnetic field [53]. When the field was horizontal and parallel to the wall, it did not affect ferrofluid spreading, whereas a horizontal field perpendicular to the wall reduced the meniscus height. On the other hand, the vertical field enhanced spreading and increased meniscus height. A radial magnetic field on a horizontal substrate can be used to induce spreading of a ferrofluid film analogously to centrifugal force used in spin coating [57]. Contrary to spin coating, the spreading pattern could be tuned by using a perpendicular field to alter the film shape before turning on the radial field. Wetting dynamics of a ferrofluid thin film was also recently investigated in more detail using image analysing interferometry [56]. When the film was exposed to a non-uniform magnetic field created by a cylindrical permanent magnet, an interesting transient increase in the ferrofluid surface curvature near the contact line region was observed during the first 2-3 seconds, followed by a small increase in the adsorbed film thickness and spreading of the entire film. The advancement of the film could be changed significantly by varying the magnetic force.

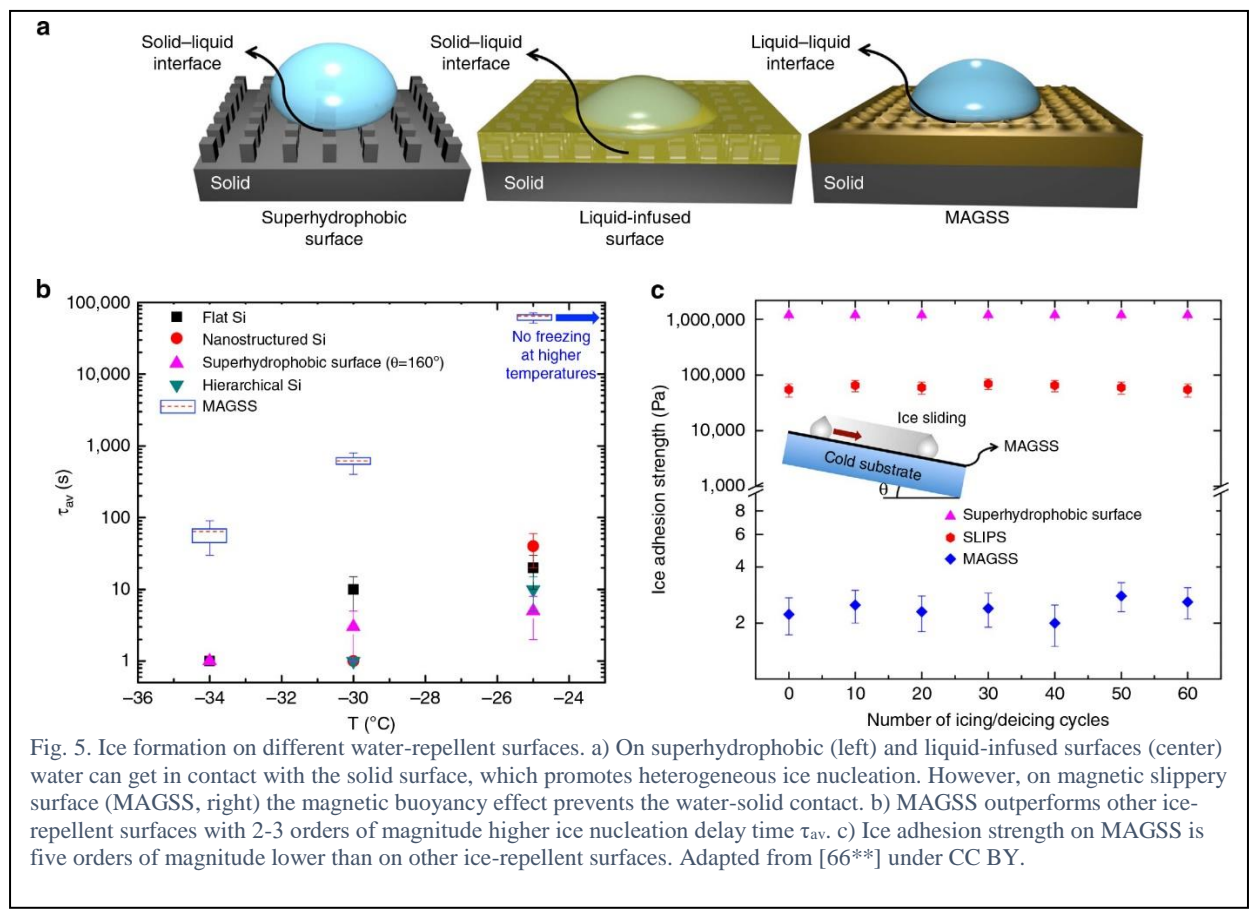

Fig. 5. Ice formation on different water-repellent surfaces. a) On superhydrophobic (left) and liquid-infused surfaces (center) water can get in contact with the solid surface, which promotes heterogeneous ice nucleation. However, on magnetic slippery surface (MAGSS, right) the magnetic buoyancy effect prevents the water-solid contact. b) MAGSS outperforms other ice-repellent surfaces with 2-3 orders of magnitude higher ice nucleation delay time $\tau_{av}$. c) Ice adhesion strength on MAGSS is five orders of magnitude lower than on other ice-repellent surfaces. Adapted from [66**] under CC BY.

## 3.2 Ferrofluids in controlling and characterizing wetting

Interplay between capillary, magnetic and other forces can lead to complex and interesting phenomena. For example, rotating ferrofluid droplets in external magnetic field undergo different kinds of instabilities depending on their wetting state [58]. Magnetic capillary origami, that is an elastic membrane wrapped around a ferrofluid droplet, can be controlled with a magnetic field and display an overturning instability when a critical magnetic field value is reached [59]. Similar magnetic control schemes can be applied to wetting phenomena.

Perhaps the most extreme example of wetting control is to prevent wetting altogether. With ferrofluid this can be realized by suspending a droplet in air with a computer-controlled magnetic field [60]. Wetting states of sessile droplets can be similarly manipulated with implications to droplet transport applications. Cheng *et al.* controlled mobility of a ferrofluid droplet on a superhydrophobic iron surface by magnetizing and demagnetizing the surface [61]. In a demagnetized state the droplet moved easily with a roll-off angle of 7°, but remained pinned on a magnetized surface. The authors speculated that this is due to a transition from a Cassie-Baxter to a Wenzel state (the former is a wetting state where some air remains trapped between the liquid and parts of the solid, the latter describes a wetting state where the liquid has completely penetrated the microscopic surface structure [15*]). They later demonstrated that a ferrofluid droplet can be reversibly switched between Cassie-Baxter and Wenzel states and showed qualitatively that a larger magnetic force is needed for Wenzel to Cassie-Baxter transition [62]. Magnetically induced pressure required for the Wenzel transition has been measured by Al-Azawi *et al*. for both static and laterally moving ferrofluid droplets on micropillared surfaces (Fig 6 a) [63*]. Moving droplets were observed to collapse more easily to the Wenzel state than static droplets. A wetting transition can also be induced with a weak dynamic magnetic field when the field frequency is near the resonant frequency of the droplet [64]. Oscillating sessile ferrofluid droplets have also been recently used for microrheological measurements [65].

Ferrofluids can also be used as a surface or a template to control wetting properties of other liquids. Peyman *et al*. recently introduced a ferrofluid-based ice-resistant surface, which could delay ice formation orders of magnitude more than state-of-the-art counterparts (Fig. 5) [66**]. They used silicon wafers coated with a 300 µm thick layer of oil-based ferrofluid, which was exposed to a magnetic field created by a permanent magnet underneath the substrate. The magnetic field pulls the ferrofluid against the substrate, creating a magnetic buoyancy effect which forces any water droplets to float near the ferrofluid surface [2]. This prevents water from touching the solid surface and impedes heterogeneous ice nucleation. On these surfaces ice formation temperature is lowered to -34 °C, approximately 10 °C lower than on superhydrophobic or slippery liquid-infused porous surfaces (SLIPS) [67]. Furthermore, the ice adhesion force is extremely low, making any formed ice easy to remove. A similar scheme was used by Khalil *et al*. to manipulate non-magnetic droplets and solid particles on ferrofluid infused micro-pillared surfaces [68]. As with other SLIPS the ferrofluid acts as a lubricant, but here it also covers the droplet or the particle, allowing it to be pulled with a permanent magnet. Recently also an array of ZnO nanorods infused with ferrofluid was used in a similar way [69*]. Capillary pressure keeps the ferrofluid between the rods and the surface remains smooth. However, the ferrofluid self-assembles in microstructures on the surface when a perpendicular magnetic field is applied. This changes the surface wetting properties, increasing the water contact angle from approximately 30° to almost 90°. The authors claim that the magnetic field gradient leads to different contact angles on opposite sides of the droplet, which can be used to drive the droplets with velocities up to 0.9 m/s. Magnetically induced roughness was used to change wetting properties also on an elastomer surface embedded with iron microparticles [70*]. The contact angle changed from approximately 100° to 165° when a 250 mT magnetic field was applied, while sliding angle decreased to 10°. Without magnetic field the droplets remained pinned on the surface even when it was turned upside down. More information about liquid transport on surfaces responsive for external fields, including magnetic fields, can be found in a recent review by Li *et al*. [71*]. Ferrofluid has also recently been indirectly used to control wetting of water droplets. Huang *et al*. used magnetic disks of 150 µm diameter to create an array of ferrofluid spikes, which were used as a master for molding arrays of polymer microcones [72]. By controlling the magnetic field direction, orientation of the spikes could be controlled and the resulting surface showed anisotropic wetting properties with 40% higher retention force for droplets moving against the direction of cone inclination than along it.

Ferrofluid can also be used for droplet manipulation without a solid substrate. Yang and Li used a thin ferrofluid film floating on a water surface to move and coalesce water droplets placed on the film [73]. In addition to controlling the droplets on the film, they could be pulled to the ceiling or the bottom of the container. The ferrofluid film on the surface of the droplet prevents the droplet from mixing with the surrounding water, unless a strong enough magnetic field is used to break the film, making a controlled release of the contents of the droplet possible.

Highly concentrated ferrofluids show spectacular instabilities and strong response to magnetic field, but also dilute magnetic liquids can be useful. A minute amount of superparamagnetic nanoparticles can render a water droplet controllable with magnetic fields without significantly altering other physical properties of water. This approach was used to measure the wetting properties of superhydrophobic surfaces by observing the motion of a water-like ferrofluid droplet in a parabolic magnetic potential [74]. The droplet was displaced from its potential energy minimum at the axis of a cylindrical permanent magnet and allowed to oscillate freely on the surface. Dissipative forces relating to contact angle hysteresis and viscous dissipation in the droplet were calculated by measuring the damping rate of the oscillations. Investigating mobility directly from a moving droplet overcomes many problems related to indirect methods such as contact angle goniometry, which suffers optical inaccuracies when measuring superhydrophobic surfaces [75]. The same method was later used to measure energy dissipation on micropillared surfaces and its dependency on solid fraction and field strength (Fig 6 b) [63*]. Droplet adhesion forces on fibers have also been characterized with ferrofluids by placing droplets on the fibers and bringing a permanent magnet incrementally closer until the droplet detaches [76]. The fibers were attached to a balance, which was used to measure the maximum magnetic force acting on the droplet before detachment.

### 3.3 Droplet actuation

Wetting plays a critical role in many applications of ferrofluids. This is obvious for example in magnetic droplet actuation, where droplet mobility depends on the surface wetting properties. Mats *et al*. investigated magnetic actuation of aqueous droplets with micrometer-sized magnetic particles on three different surfaces: PTFE, superhydrophobic *Colocasia* leaf and glass slide coated with a commercial superhydrophobic spray solution [77]. The commercial spray proved to be best suited for magnetic actuation, while PTFE showed higher friction and the leaf was damaged by the magnetic microparticles. It is noteworthy that the static contact angles on PTFE were almost independent of the microparticle concentration, but decreased over 20 degrees on the superhydrophobic surfaces as concentration was increased. This was attributed to microparticles becoming associated with the surface and covering the micro- and nanostructures necessary for superhydrophobicity. Furthermore, an applied magnetic field reduced the static contact angle another 20 degrees on the *Colocasia* leaf, while the effect was insignificant on PTFE and glass coated with superhydrophobic spray. This is probably because *Colocasia* has microstructures at the same scale as the magnetic particles, unlike the two other surfaces. This goes to show that a high static water contact angle does not guarantee that the surface is suitable for actuation of aqueous magnetic droplets.

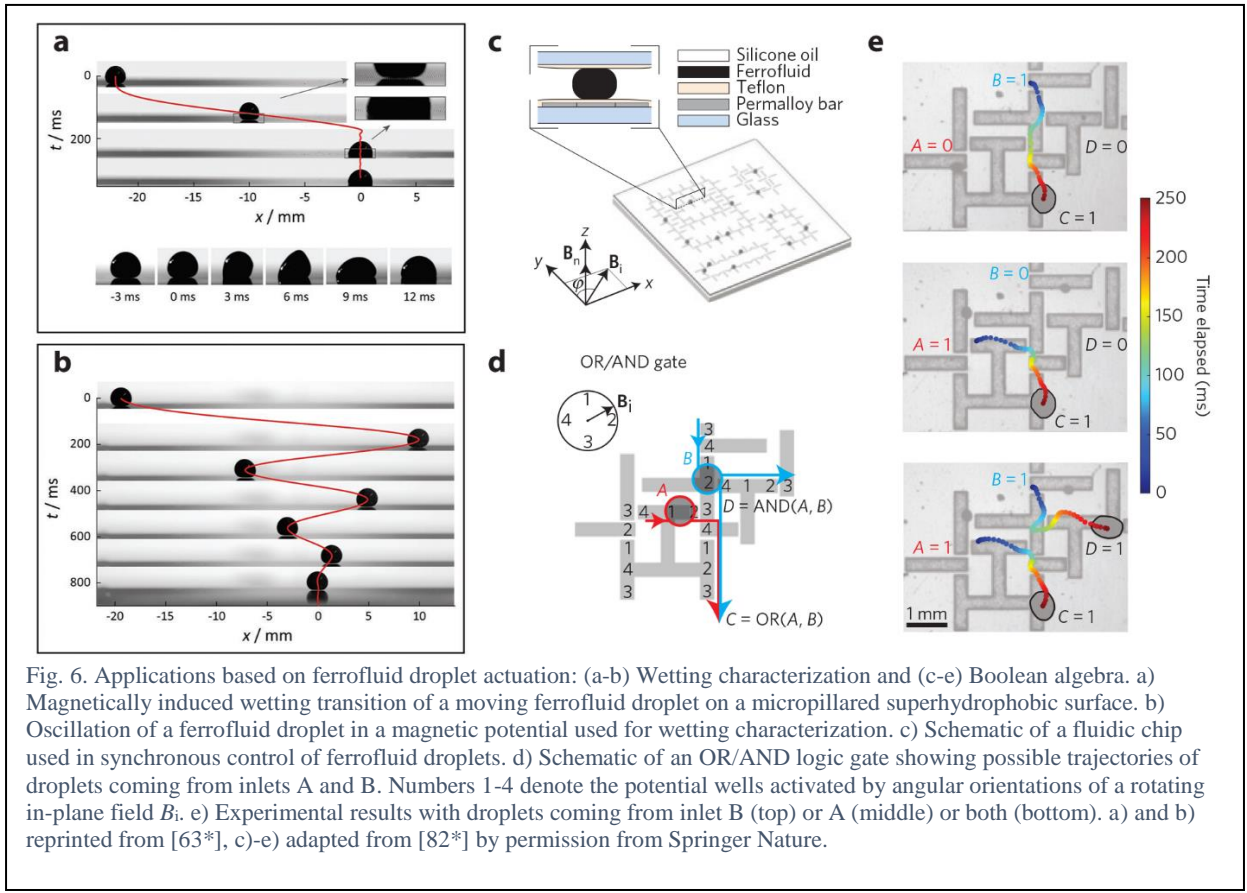

Fig. 6. Applications based on ferrofluid droplet actuation: (a-b) Wetting characterization and (c-e) Boolean algebra. a) Magnetically induced wetting transition of a moving ferrofluid droplet on a micropillared superhydrophobic surface. b) Oscillation of a ferrofluid droplet in a magnetic potential used for wetting characterization. c) Schematic of a fluidic chip used in synchronous control of ferrofluid droplets. d) Schematic of an OR/AND logic gate showing possible trajectories of droplets coming from inlets A and B. Numbers 1-4 denote the potential wells activated by angular orientations of a rotating in-plane field $B_i$. e) Experimental results with droplets coming from inlet B (top) or A (middle) or both (bottom). a) and b) reprinted from [63*], c)-e) adapted from [82*] by permission from Springer Nature.

Ferrofluids have been increasingly studied and used in continuous-flow microfluidics applications [11*], [12]. Here magnetic fields allow control over droplet generation, coalescence and separation, as has been recently investigated both experimentally and numerically [78]–[80]. On the other hand, digital microfluidics and manipulation of individual droplets has so far been dominated by electric actuation. Magnetic actuation in this context has mostly been realized using large magnetic particles [13*], [14*]. However, progress is also being made in digital microfluidics using ferrofluids. Chakrabarty *et al*. presented a numerical analysis of magnetic manipulation of a ferrofluid droplet using a micro-coil array, which could be used to perform complex droplet functions [81]. Katsikis *et al*. introduced a platform for synchronous droplet transport capable of basic droplet-based computation [82*]. The system consists of aqueous ferrofluid droplets immersed in silicone oil between two glass plates (Fig 6 c). One of the plates is coated with permalloy patterns, which are magnetized with an in-plane rotating field. As the field rotates, the magnetization of the patterns changes, which induces a synchronous motion of the droplets. The droplets are also magnetized with an out-of-plane magnetic field, which induces magnetic repulsion between them. This allows designing logic gates, where droplet trajectories depend on other droplets in the system (Fig 6 d and e).

## 4. Conclusions

Wetting of ferrofluids is affected by a number of phenomena. Surfactants used in nanoparticle stabilization tend to lower the surface tension compared to pure carrier liquid, whereas nanoparticle adhesion to the solid surface can cause complex and counterintuitive dynamic wetting phenomena. Magnetic fields induce body forces and surface stresses on ferrofluid droplets, which can cause either droplet flattening or elongation depending on the field geometry. These deformations also affect apparent contact angles, even if intrinsic wetting properties remain unchanged. In addition, there is evidence that magnetic fields can slightly affect interfacial tensions of ferrofluids via nanoparticle dipole interactions and changing particle distribution within the fluid.

When investigating ferrofluid wetting properties using contact angle goniometry, all these different contributions must be carefully considered. High enough magnification must be used to distinguish field-induced curvature of a ferrofluid surface from the contact angle determined by the wetting properties. Both advancing and receding contact angles should be measured to avoid confusing a change in intrinsic wetting properties with a flow-induced change in the contact angle. Previous studies have often reported only a single static contact angle value, which is not enough to accurately describe the wetting system.

To better understand and control ferrofluid wetting, further systematic, high quality experimental studies are needed. This will facilitate development of magnetic droplet transport and other applications relying on ferrofluid actuation.

## Acknowledgements

This work was supported by the European Research Council ERC-2016-CoG (725513-SuperRepel) and the Academy of Finland (Centres of Excellence Programme (2014−2019) and Academy Postdoctoral Researcher).